\documentclass{kapproc} 
\usepackage{epsfig} 
\def\apj{ApJ}                                   
\def\apjl{ApJ}                                  
\def\apjs{ApJ Supp.}                            
\def\mnras{MNRAS}                               
\def\aap{Astron. Astr.}                         
\def\aapl{Astron. Astr. (Lett.)}                
\def\apss{Adv. Space Sci.}                      
\def\nat{Nature}                                
\def\prd{Phys. Rev. D}                          
\def\ssr{Space Sci. Rev.}                       
\def\teq#1{$\, #1\,$}                           
{\catcode`\@=11                                                   
\gdef\SchlangeUnter#1#2{\lower2pt\vbox{\baselineskip 0pt\lineskip0pt     
\ialign{$\m@th#1\hfil##\hfil$\crcr#2\crcr\sim\crcr}}}}            
\def\gtrsim{\mathrel{\mathpalette\SchlangeUnter>}}                
\def\lesssim{\mathrel{\mathpalette\SchlangeUnter<}}  

\font\sixrm=cmr6 scaled \magstep0 
\def\dover#1#2{\hbox{${{\displaystyle#1 \vphantom{(} }\over{ 
   \displaystyle #2 \vphantom{(} }}$}}         
\def\erg{\varepsilon}     
\def\lambar{\lambda\llap {--}_{\rm c}} 
\def\emax{\erg_{\hbox{\sixrm MAX}}} 
\def\emin{\erg_{\hbox{\sixrm MIN}}} 
\def\gammamin{\gamma_{\hbox{\sixrm MIN}}} 
\def\dpsr{\hbox{$d_{\hbox{\sixrm PSR}}$}} 
\def\dpsrsq{\hbox{$d^2_{\hbox{\sixrm PSR}}$}} 
\def\ethresh{\erg_{\hbox{\sixrm TH}}} 
\begin{document} 
 
\newcommand{\vol}[2]{$\,$\bf #1\rm , #2.}                  
 
\articletitle[Polar Cap Pulsar Models] 
{Identifying the Mysterious\\ 
 EGRET Sources: Signatures\\ 
 of Polar Cap Pulsar Models} 

\vphantom{p}
\vskip -145pt
\centerline{\hfill To appear in Proc. Tonantzintla Workshop {\it The Nature of}}
\centerline{\hfill {\it Unidentified Galactic High-Energy Gamma-Ray Sources}}
\centerline{\hfill eds. A. Carrami\~nana, et al.
             (Kluwer Academic, Dordrecht, 2001)}
\vskip 100pt

\author{Matthew G. Baring} 
\affil{Department of Physics and Astronomy,\\ 
Rice University, MS-108,\\  
P. O. Box 1892,\\ 
Houston, TX 77251-1892,\\ 
U. S. A.} 
\email{baring@rice.edu} 
 
 
\begin{abstract} 
The advent of the next generation of gamma-ray experiments, led by 
GLAST, AGILE, INTEGRAL and a host of atmospheric \v{C}erenkov telescopes 
coming on line in the next few years, will enable ground-breaking 
discoveries relating to the presently enigmatic set of EGRET/CGRO UID 
galactic sources that have yet to find definitive identifications. 
Pulsars are principal candidates for such sources, and many are 
expected to be detected by GLAST, some that are radio-selected, like 
most of the present EGRET/Comptel pulsars, and perhaps even more that 
are detected via independent pulsation searches.  At this juncture, it 
is salient to outline the principal predictions of pulsar models that 
might aid identification of gamma-ray sources, and moreover propel 
subsequent interpretation of their properties.  This review summarizes 
relevant characteristics of the polar cap model, emphasizing where 
possible distinctions from the competing outer gap model.  Foremost 
among these considerations are the hard X-ray to gamma-ray spectral 
shape, high energy cutoffs and pulse profiles, and how these 
characteristics generally depend on pulsar period and period 
derivative, as well as observational viewing angle.  The polar cap 
model exhibits definitive signatures that will be readily tested by the 
detections of GLAST and other experiments, thereby establishing cogent 
observational diagnostics.  The paper focuses on different classes of 
pulsars that might define agendas and parameter regimes for blind 
gamma-ray pulsation searches; examples include the highly-magnetized 
ones that are currently quite topical in astrophysics. 
\end{abstract}

\section{Introduction} 
Pulsars are a central part of any discussion of candidates for the 
putatively galactic population of EGRET unidentified (UID) sources.  A 
major factor in this is their inherent nature in being both among the 
brightest galactic sources (for observational summaries, see Kanbach, 
these proceedings, Thompson 2001), and moreover being distinctive via 
their pulsations.  These features have fed the historical evolution of 
the field of unidentified gamma-ray sources, with pulsars leading the 
{\it post-facto} galactic identifications (e.g. see Thompson, these 
proceedings).  There is a pervasive feeling in the gamma-ray community, 
as embodied in the course of this Workshop, that such a situation will 
persist.  This perception is driven by the expectation that the 
Gamma-Ray Large Area Space Telescope (GLAST:  {\tt 
http://www-glast.stanford.edu}) will detect pulsars in profusion, some 
that are radio-selected, like most of the present EGRET/Comptel 
pulsars, and perhaps even more that are detected via independent 
(blind: i.e. not radio or X-ray selected) pulsation searches.  Current 
estimates of the anticipated GLAST pool of pulsars range from dozens to 
several hundred (e.g. Harding, 2001b), depending on whether an outer 
gap or polar cap model is preferred, and on specific assumptions 
pertaining to each model.  This population should account for a 
significant fraction of the present EGRET UID galactic sources. 
Furthermore, it should provide an ample dataset for exploring pulsar 
physics in general, and enable discrimination between the outer gap and 
polar cap scenarios in particular. 
 
Discussions of the generic features of each leading model for gamma-ray 
emission in pulsars are therefore timely.  The outer gap (OG) case, 
where the accelerating potential is far removed from the stellar 
surface and more proximate to the light cylinder, is advocated by 
Romani (these proceedings).  This paper presents the perspective of the 
polar cap (PC) model, where particle acceleration is effected either 
just above or within a stellar radius or two of the neutron star 
surface, and is presumed to occur on the open field lines
near the polar cap (e.g. Sturrock 1971; Ruderman \& Sutherland 1975; 
Arons \& Scharlemann 1979; see also the recent review by Harding 
2001a).  Motivations for preferring a polar cap scenario include, but 
are not limited to, (i) a difficulty in accepting a viewpoint that 
electrodynamic dissipation near the stellar surface is only a minor 
contributor to the pulsar's high energy radiative signals, (ii) the 
connection of gamma-rays to coherent radio emission in PC models, 
mediated by single photon pair creation in strong magnetic fields, 
(iii) the relatively narrow range of ages permitted for gamma-ray 
pulsars according to the outer gap model, (iv) the fact that the polar 
cap prediction (Harding 1981) of pulsar luminosity scaling 
approximately with open line field voltage was confirmed by the 
EGRET/Comptel pulsar collection (e.g. see Fig.~4 of Thompson 2001, for 
the status quo of this correlation), and (v) the predictions (Cheng, Ho 
\& Ruderman 1986; Cheng 1994; Yadigaroglu \& Romani 1995; Romani 1996) 
from earlier inceptions of outer gap models of many radio-quiet 
gamma-ray pulsars and also inverse Compton emission at TeV energies, 
forecasts that have not been borne out by subsequent observations. 
While the first two of these are largely conceptual preferences, the 
last two connect to reality.  Geminga stands alone is the only 
potentially radio-quiet ``garden-variety'' gamma-ray pulsar, and the 
constraining limits to TeV (pulsed) emission from pulsars obtained by 
Atmospheric \v{C}erenkov Telescopes (e.g. Nel et al. 1993; Lessard et 
al. 2000) have forced the revision (e.g. Hirotani, 2000) of TeV flux 
estimates from outer gap models. 
 
The features of the polar cap model are intimately connected to the 
extremely strong magnetic field that threads the emission region.  It 
is this field that controls the maximum energy of emission, the nature 
of pair creation, and a host of physics that determines the pulsar 
spectrum and influences the pulse shape.  This review will highlight 
some of these properties, and explore how (mostly) spectral and temporal 
characteristics generally depend on pulsar period and period 
derivative, as well as on observational viewing angle.  The polar cap 
model exhibits definitive signatures that will be readily tested by the 
detections of GLAST and other experiments, thereby establishing 
palpable observational diagnostics.  Discerning which of the outer gap and  
polar cap models is most appropriate for gamma-ray pulsars, or whether 
each has its own domain of applicability, is a central quest for 
pulsar astrophysicists.  While preparatory analysis for the 
GLAST era is a worthwhile goal alone, perhaps most salient for the 
subject of this meeting is a definition of identification strategies 
for gamma-ray sources, based on the global characteristics of pulsar 
models and the assumption that pulsars constitute a sizeable fraction 
of the EGRET UID collection; this will form the focus at the end 
of this review.

\section{Polar Cap Models of Gamma-Ray Pulsars} 
 
The polar cap scenario is attractive from a physics perspective. 
Nevertheless, like its outer gap competitor, it remains unproven. 
Hence it is imperative to establish definitive/unambiguous properties and 
predictions that enable a determination of its applicability to 
gamma-ray pulsars.  This can be achieved via two approaches.  The first 
is to isolate individual pulsars for analysis, and explore the model 
behavior of phase-resolved spectra (and also polarization swing 
profiles in an ideal world where gamma-ray polarimetry is accessible: 
see Section~\ref{sec:pcpred}) that can be compared to high quality 
temporal and spectral data.  This has been the approach of Daugherty \& 
Harding (1996), Dyks \& Rudak (2000) and Romani (1996) using the Vela 
pulsar as a test case in supporting their competing perspectives.  The 
major drawback of such isolated focuses is that the models have enough 
parameters to render model discrimination near impossible.  Therefore, 
multiple objects need to be considered, a time-consuming task for 
pulsar-by-pulsar analyses.  This leads to the second diagnostics 
method: population statistics.  It is expedient to identify global 
characteristics of the models so as to define parametric trends that 
can be confirmed or disproven given a large database such as that to be 
afforded by GLAST.  Given the disparity in emission region geometry and 
operable physics incorporated in the polar cap and outer gap scenarios, 
it is improbable that they will provide a collection of similar or 
coincident behavioral trends.  Hence, the emphasis of this paper will 
be to address the more global signatures of polar cap models by first 
defining the relevant cascade and radiative properties. 
 
\subsubsection{Basic Properties of Cascades} 
\vskip 10pt 
 
Polar cap models for pulsar high-energy emission are generally based on 
the idea, dating from the earliest pulsar models of Sturrock (1971) and 
Ruderman \& Sutherland (1975; hereafter RS75), of particle acceleration 
and radiation near the neutron star surface at the magnetic poles. 
Within this broad class, there is a large variation, with the primary 
division being whether or not there is free emission of particles from 
the neutron star surface.  This question hinges on whether the surface 
temperature \teq{T} of the neutron star (many of which have now been 
measured in the range \teq{T \sim 10^5 - 10^6} K; Becker \& Tr\"umper 
1997) exceeds the ion, \teq{T_{\rm i}} and electron, \teq{T_{\rm e}}, 
thermal emission temperatures.  If \teq{T < T_{\rm i}}, a vacuum gap 
will develop at the surface, due to the trapping of ions in the neutron 
star crust (RS75, Usov \& Melrose 1995).  In this case, the particle 
acceleration and radiation will take place very near the neutron star 
surface.  If \teq{T > T_{\rm e}}, free emission of particles of either 
sign of charge will occur.  The flow of particles is then limited only 
by space charge, and an accelerating potential will develop (Arons \& 
Scharlemann 1979; Muslimov \& Tsygan 1992) due to an inability of the 
particle flow all along each open field line to supply the corotation 
charge that is required to short out the electric field component 
\teq{E_{\parallel}} along the magnetic field lines.  In space 
charge-limited flow models, the accelerating \teq{E_{\parallel}} is 
screened at a height where the particles radiate \teq{\gamma}-rays that 
produce pairs.  This so-called pair formation front (e.g.  Arons 1983, 
Harding \& Muslimov 1998) can occur at high altitudes above the polar 
cap, a property that may prove necessary to explain the spectral 
cutoffs in the some or most of the EGRET pulsars. 
 
The acceleration of primary electrons is rapid and ceases when one of 
two types of radiative cooling becomes significant.  This establishes 
the maximum Lorentz factor \teq{\gamma_e} of these particles, and a 
quasi-monenergetic primary distribution is established prior to 
cascading.  The two cooling mechanisms are curvature radiation induced 
by the magnetic field line curvature, the process that is more widely 
cited in pulsar literature as a primary emission mechanism, and 
resonant (magnetic) inverse Compton scattering of thermal X-rays from 
the stellar surface (e.g. Sturner and Dermer 1994), a relatively recent 
consideration.  Both are strong functions of the magnetic field 
strength and either the electron's Lorentz factor or the field 
geometry.  Curvature radiation-initiated cascades generally have 
\teq{\gamma_e\sim 10^7} (e.g. Daugherty \& Harding 1989; see also 
Harding \& Muslimov 1998), while inverse-Compton seeded pair cascades 
yield \teq{\gamma_e\sim 3\times 10^5}--\teq{10^6} (e.g.  Sturner 1995; 
see also Harding \& Muslimov 1998).  Such photons propagate through the 
magnetosphere until they achieve sufficient angles \teq{\theta_{\rm 
kB}} with respect to the magnetic field to permit the creation of pairs 
via \teq{\gamma\to e^+e^-} above the threshold energy of 
\teq{2m_ec^2/\sin\theta_{\rm kB}}.  This propagation is influenced by 
general relativistic distortions of photon trajectories and field 
structure (e.g. Gonthier \& Harding 1994; Harding, Baring \& Gonthier 
1997), as is the magnitude of the field in the local inertial frame. 
For small polar cap sizes, corresponding to longer pulsar periods, it 
is the failure of the primary photons to acquire sufficient angles 
\teq{\theta_{\rm kB}} at low to moderate altitudes (prior to dipole 
field decline) that is primarily responsible for the existence of a 
theoretical death line for radio pulsars (Sturrock, Baker \& Turk 
1976): pair creation is quenched at high altitudes since the rate is a 
strongly increasing function of {\bf B} (e.g. Tsai \& Erber, 1974). 
 
The first generation of pair creation initiates the pair cascade, with 
pairs generally being created in excited transverse (to the field) 
momentum states, the so-called Landau levels.  De-excitation via 
cyclotron and synchrotron radiation is then extremely rapid, on 
timescales of \teq{10^{-16}}sec or less for typical neutron star fields 
of \teq{B_0\gtrsim 10^{12}}Gauss (n.b. subscripts zero denote polar 
surface fields).  These secondary photons can then travel to higher 
altitudes and create further pairs and successive generations of 
photons.  So proceeds the cascade, with a typical number of generations 
being around 3--4, and the total number of pairs per primary electron 
being of the order of \teq{10^3}--\teq{10^4} (Daugherty \& Harding 
1982).  The cumulative product is an emission spectrum that comprises a 
curvature/inverse Compton continuum that is cut off at hard gamma-ray 
energies by pair creation, with the addition of several synchrotron 
components at successively lower energies, terminating when the 
magnetosphere becomes transparent to \teq{\gamma\to e^+e^-} at 
significant altitudes.  The details of such spectral formation are 
discussed below.  A notable exception to this cascade scenario arises 
in highly-magnetized pulsars, PSR 1509-58 (with \teq{B_0\sim 3\times 
10^{13}}Gauss) being the case in point.  When the surface polar field 
\teq{B_0} exceeds around \teq{6\times 10^{12}}Gauss, pairs are produced 
in the zeroth (ground state) Landau level (Baring \& Harding 2001, 
hereafter BH01), so that cyclotron/synchrotron emission is prohibited. 
Cascading is then effectively squelched and the pair yield diminished 
(BH01).  A possible amelioration of this circumstance was posited by 
Zhang \& Harding (2000a), namely the Landau level excitation of higher 
generation pairs via Compton scatterings with X-rays from the surface. 
Since such excitation can only arise above the cyclotron resonance 
(e.g. Gonthier et al. 2000), Baring \& Harding (2001) determined that 
the population of excited Landau states relative to that in the ground 
state is small. 
 
A more fascinating variation on the cascade theme involves the 
phenomenon of photon splitting, again applicable to high field 
pulsars.  Magnetic photon splitting, \teq{\gamma\to\gamma\gamma}, a 
third-order (in the fine structure constant \teq{e^2/\hbar c}) quantum 
electrodynamical (QED) process in which a single photon splits into two 
lower-energy photons (Adler 1971, Baring \& Harding 1997), operates 
efficiently and competes effectively with pulsar pair production only 
in magnetic fields above \teq{\sim 10^{13}}Gauss (Harding, Baring \& 
Gonthier 1997, hereafter HBG97).  This region of high magnetic field 
strength lies in the upper-right part of the \teq{P-\dot P} diagram. 
Splitting is forbidden in field-free regions by Furry's theorem, a 
symmetry property of QED.  In regimes of weak or modest fields when 
vacuum dispersion effects are small, \teq{\gamma\to\gamma\gamma} is a 
collinear process, conserving both energy and momentum.  The rate of 
splitting, like that of magnetic pair creation \teq{\gamma\to e^+e^-}, 
is generally a rapidly increasing function of field strength (the exception 
being at fields \teq{B\gtrsim 10^{14}}Gauss), photon energy and photon 
propagation angle with respect to the field.  However, splitting 
possesses no energy threshold, so that it can and does dominate the 
first order process of pair creation if {\bf B} is sufficiently high. 
This leads to an alternative channel for cascade cessation, with 
gamma-rays being reprocessed without yielding pairs so that synchrotron 
generations are suppressed.  The result is distinctive bumps and 
polarization signals in the EGRET/Comptel band (HBG97).  The issue of 
splitting-influenced pulsar cascades is addressed below in 
Section~\ref{sec:pcpred} and in depth in HBG97 and Baring \& Harding (2001). 
 
\section{Predictions of Polar Cap Models} 
 \label{sec:pcpred} 
 
Having assembled the basic ingredients of pulsar cascades, the task 
here is to identify the array of radiative signatures that are the 
hallmark of polar cap models.  Since the cascade physics differs 
according to the period \teq{P} and period-derivative \teq{{\dot P}} of 
the pulsar, for the purposes of this discussion a division is made 
between the canonical Crab-like/Vela-like young pulsars with moderately 
strong fields, and their more highly-magnetized cousins like PSR 
1509-58.  The focus in this section is mainly on pulsars whose beam 
sweeps across the line of sight to Earth, so-called {\it on-beam} pulsars; 
off-beam pulsars will also be mentioned briefly. 
 
\subsubsection{Crab-like and Vela-like pulsars} 
 \label{sec:Velalike} 
 
Here we consider ``standard'' bright young pulsars like the Crab, Vela 
and Geminga with moderately high, but not extremely high, surface polar 
fields \teq{B_0}.  These constitute the majority of EGRET/Comptel 
pulsars, with PSR 1509-58 (addressed in the next subsection) and the 
millisecond pulsar PSR J0218+4232 (Kuiper et al. 2000) being the 
notable exceptions.  Spectral properties will be the first focus.  From 
the discussion of cascade properties above, this case corresponds to 
synchrotron-curvature cascades if the pulsar period is substantially 
shorter than a second.  The curvature spectrum is generated by a 
quasi-monoenergetic electron injection that results from the rapid 
electrodynamic acceleration.  Since the curvature mechanism is 
essentially identical to the synchrotron one, except that the relevant 
curvature scale is the radius of field curvature as opposed to the 
gyroradius, simple synchrotron formalism can be applied (e.g. see 
Jackson 1975) to yield a spectrum of \teq{\erg^{-2/3}} below a maximum 
cutoff energy.  This result holds provided that there is no cooling 
during emission, which is generally not the case.  With curvature 
cooling operating (i.e. when \teq{P\lesssim 0.3}sec), the spectrum 
steepens to \teq{\erg^{-5/3}} (e.g. Daugherty \& Harding 1982), a 
power-law that extends down to energies at which cooling becomes 
inefficient, and then the flat \teq{\erg^{-2/3}} form is again 
assumed. 
 
Superposed on this are contributions from synchrotron emission from 
successive generations.  Since these components result from pairs that 
are created by the photon spectrum belonging to a previous cascade 
generation, and the pair injection traces the input photon spectrum, it 
is straightforward (e.g. Wei, Song \& Lu 1997; Harding \& Daugherty 
1998; Baring \& Harding 2000) to determine that the spectral index 
\teq{\alpha_i} of the \teq{i^{th}} generation (i.e. for 
\teq{dn_\gamma^{(i)}(\erg )/d\erg\propto \erg^{-\alpha_i}}) satisfies 
the recurrence relation \teq{\alpha_{i+1}=(\alpha_i +1)/2}.  This 
result assumes that synchrotron cooling of the pairs rapidly 
depopulates excited Landau levels and steepens the pair spectrum by an 
index of unity.  It then follows that (Harding \& Daugherty 1998) 
\begin{equation} 
\alpha_n = 2 - \dover{2-\alpha_1}{2^{n-1}}\quad , 
 \label{eq:alpha_n} 
\end{equation} 
where the primary index is \teq{\alpha_1\approx 5/3}.  Hence, each 
generation steepens in spectral index, so that, due to its 
preponderance of photons, the last generation determines the emergent 
cascade index.  Notice that the asymptotic index for a large number of 
generations is 2.  This automatically creates difficulties within this 
framework for pulsars with spectra steeper than \teq{\erg^{-2}}, the 
Crab being the notable case.  However, it must be remembered that such 
simple analytic determinations are ``one-zone'' computations, and that 
the cascade spans a range of altitudes and field geometries, all of 
which modify this picture of steepening.  In particular, the finite 
generational energy degradation \teq{\chi =0.5 \max [B/B_{\rm cr}, \; 
0.1]} (i.e.  such that photon energies satisfy the recurrence relation 
\teq{\erg_{i+1}\sim\chi\erg_i}) disrupts the idealized picture of 
power-laws generating power-laws, since the structure introduced by 
pair creation cutoffs (discussed below) influences successive 
generations to introduce additional steepening.  Note that, hereafter, 
\teq{B_{\rm cr}=4.413\times 10^{13}}G denotes the quantum critical 
field. 
 
The spectral index is not a free parameter, but is determined roughly 
by noting that cascade cessation occurs when the mean free path for 
pair creation is comparable to the stellar radius \teq{R_0}, i.e. the 
scalelength for field decline.  This then establishes (e.g. Harding, 
private communication; Baring \& Harding 2000) the effective maximum 
number of cascade generations (permitted to be a non-integer, following 
Lu, Wei \& Song 1994) 
\begin{equation} 
n = 2 + \dover{1}{\log_e\chi}\, \log_e\Biggl[ 
\dover{4}{3\pi}\, \dover{B_{\rm cr}}{B_0\gamma_1^3}\, 
\biggl( \dover{r}{R_0} \biggr)^4\, \dover{Pc}{\lambar}\, \Biggr] 
 \label{eq:gener} 
\end{equation} 
for pulsar period \teq{P} (in seconds) and electron primary Lorentz 
factor \teq{\gamma_1}.  Here \teq{\lambar} is the Compton wavelength 
over \teq{2\pi}.  Observe the appearance of a local field \teq{B_0 
(r/R_0)^{-3}} factor.  This relation provides a closed system, 
establishing the spectral index as a function of spin-down parameters 
\teq{P} and \teq{B_0} and the (unknown) altitude \teq{r-R_0} of typical 
emission above the surface.  Variations on this {\it generation index} 
formulation can be found in Lu, Wei \& Song (1994) and Wei, Lu \& Song 
(1997), and also in Zhang \& Harding (2000a) who treat (resonant) 
inverse Compton cascading as well to probe spectral signatures in the X-ray 
band.  While details of ``altitudinal smearing'' will muddy this 
spectral index determination (as is evident in the comparison of such 
predictions with EGRET pulsar characteristics in Harding \& Daugherty 
1998), one expects the overall global trends with \teq{B_0} and 
pulsar period to be approximately true for polar cap models, providing 
a powerful observational diagnostic.  Generally, the number of 
generations and the spectral index are {\it increasing functions of 
\teq{B_0}, but decreasing functions of the altitude.} 
 
It must be emphasized that the mean spectral index of EGRET UIDS that
are associated with the Gould Belt is around \teq{2.25} (Grenier et
al.  2000; see also Gehrels et al. 2000), greater than that of most
EGRET pulsars; resolution of this mismatch may be found in the
discussion of off-beam pulsars below.  A brief comment on the
phase-dependence of spectra is also warranted, since pulse-phase
spectroscopy is an objective of pulsar studies, attainable for a large
number of pulsars in the GLAST era.  Constraints on model phase space
are narrowed considerably by exploration of the spectral variations
with phase, though both polar cap (Daugherty \& Harding 1996; Dyks \&
Rudak 2000) and outer gap (Romani 1996) models have successfully
accounted for Vela's properties (see also Zhang \& Cheng 2001 for an
outer gap consideration of Geminga).  A characteristic that has emerged
in gamma-ray pulsars is that their trailing peak and interpulse spectra
are flatter (e.g. Thompson 2001; see also Kanbach et al. 1994 for Vela
details) than their leading peak spectra, features that need to be
modelled.  The interpulse emission is generally expected to be harder
in the polar cap model, since there is less cascading near the pole
(Daugherty \& Harding 1996), and the spectrum possesses mostly primary
curvature radiation.  The range of altitudes probed is a function of
geometrical perspective, and hence also of pulse phase.  Given the
number of free parameters in models, it is critical to survey an array
of pulsars in model/data comparisons of spectral and temporal
properties.
 
\begin{figure}[ht] 
\vspace{-30pt} 
\centerline{\hskip 1.6truecm\epsfig{file=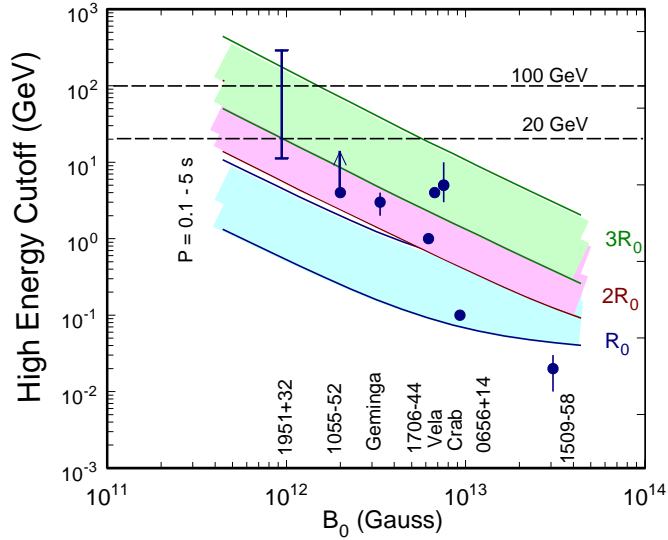, width=1.0\hsize}} 
\vspace{-20pt} 
\caption{
Maximum pulsar emission energies (from Baring \& Harding 2000) imposed 
by pair creation attenuation at different altitudes, described 
empirically via Equation~(\ref{eq:emax}).  For each altitude, a range 
of pulse periods (polar cap sizes) is represented by a shaded band. 
These energies are determined by the more involved photon 
propagation/attenuation code described in Baring \& Harding (2001). 
Inferred cutoff energies (or ranges) for 8 gamma-ray pulsars of 
different \teq{B_0} are indicated, from which a trend of declining 
altitude of emission with increasing \teq{B_0} is suggested. 
} 
\label{fig:hecutoff} 
\end{figure}

The next major spectral feature is the maximum energy of emission, 
which is controlled by attenuation due to pair creation during photon 
propagation through the pulsar magnetosphere.  This attenuation leads 
to the reprocessing that spawns lower energy photons, and provides a 
characteristic super-exponential turnover (e.g. Daugherty \& Harding 
1996) that contrasts that expected in outer gap models (e.g. see 
Thompson 2001 for a comparison). \teq{\gamma\to e^{\pm}} occurs at the 
threshold \teq{\erg\sin\theta_{\rm kB} =2} for \teq{B\gtrsim 0.1 B_{\rm 
cr}} and above threshold at \teq{\erg\sin\theta_{\rm kB} \sim 0.2 
B_{\rm cr}/B} for lower fields (e.g. see Daugherty \& Harding 1983). 
Here, \teq{\theta_{\rm kB}} is the angle of photon propagation relative 
to {\bf B}, and hereafter photon energies \teq{\erg} are expressed in 
units of \teq{m_ec^2}.  Hence, the mean free path for photon 
attenuation in {\it curved} fields is \teq{\lambda_{\rm pp} \sim 
\rho_c/\erg\, \max \{ 2, \; 0.2/B \} }, i.e.  when 
\teq{\erg\sin\theta_{\rm kB}} crosses above threshold.  The radius of 
field curvature is \teq{\rho_c = [Prc/2\pi]^{1/2}} for a pulsar period 
\teq{P}.  Pair creation cutoff energies \teq{\emax}, derived from the 
codes developed in HBG97 and Baring \& Harding (2001), are plotted in 
Figure~\ref{fig:hecutoff}.  While these are refined estimates, 
including the effects of general relativity on spacetime curvature, 
field enhancement and photon energy, their empirical dependence on 
\teq{B_0}, \teq{R_0} and pulsar period \teq{P} (in seconds) can be 
summarized in the relation (see also Harding, 2001a) 
\begin{equation} 
\emax \approx 0.4 \sqrt{P} \, \biggl( \dover{r}{R_0} \biggr)^{1/2} \; 
\max \Biggl\{ 1,\; \dover{0.1\, B_{\rm cr}}{B_0}\,  
\biggl( \dover{r}{R_0} \biggr)^3 \Biggr\}\; \hbox{GeV} \;\; . 
 \label{eq:emax} 
\end{equation} 
Refinements to this estimate to include the effects of photon splitting
in higher fields (near \teq{B_{\rm cr}}) are discussed in Baring \&
Harding (2001).  The overall trend is clear:  there is a strong
anti-correlation between the maximum energy and the surface magnetic
field, which seems to be augmented by an apparent decline of emission
altitude with \teq{B_0}.  Such a trend is a distinctive characteristic
that can be probed by GLAST and is unlikely to be reproduced by outer
gap models.  Note also, that the maximum energy is generally in the
1--10 GeV band for normal young pulsars, can be much lower (e.g. HBG97,
BH01) for highly magnetized ones, and also much higher for millisecond
pulsars (Bulik, Rudak \& Dyks 2000) so that sub TeV-band (i.e.
\teq{\sim 50}--100 GeV) signals are possible for polar cap models via
synchrotron/curvature cascades if the field is low enough.  It also
should be remarked that the cutoff energy depends on pulse phase, with
slightly greater values achieved between the pulse peaks in the case of
Vela modelling (Daugherty \& Harding 1996); such a property matches the
EGRET observations (Kanbach et al.  1994).
 
The other major spectral feature in the gamma-ray band corresponds to 
the lower energy of the cascade, which from Monte Carlo simulations 
(e.g. Daugherty \& Harding 1982) corresponds to Lorentz factors 
\teq{\gammamin} of around 50--100 for Vela-like pulsars.  The 
synchrotron photon energy for this Lorentz factor is 
\teq{\emin\sim\gammamin B/B_{\rm cr}}, noting that a factor of 
\teq{1/\gammamin} is introduced to account for the cascade beaming. 
This energy is typically in soft gamma-rays, and generally 
\teq{\gammamin} depends on \teq{B_0}, \teq{P}, etc. in more or less the 
same manner (Baring \& Harding 2000) that \teq{\emax} does in 
Eq.~(\ref{eq:emax}), since the same pair creation physics applies to 
both.  However, the reprocessing that leads to the establishment of 
\teq{\emin} occurs at lower altitudes (yielding possibly different 
parametric dependences) than the ultimate attenuation that defines 
\teq{\emax}, so that in general \teq{\emin\ll\emax}; it is difficult to 
be more specific than this inequality.  In addition, since there is 
probably little altitudinal dependence in \teq{\emin}, contrasting that 
for \teq{\emax} inferred from Figure~\ref{fig:hecutoff}, one might 
expect a spectral ``narrowing'' with increasing \teq{B_0}.  The 
spectrum possesses a break at \teq{\emin}, below which it assumes the 
flat \teq{\erg^{-2/3}} form that signifies curvature or synchrotron 
emission from quasi-monoenergetic pairs.  Such a slope is consistent 
with the broad-band optical/ hard X-ray non-thermal continuum of the 
Vela pulsar that lies underneath the thermal surface X-rays (e.g. see 
Pavlov et al.  2001 for a spectrum).  Note that when the surface field 
becomes sufficiently high, the cyclotron scale imposes structure on the 
spectrum and redefines the variation of \teq{\emin} with \teq{B_0}; 
this situation is sampled by the highly-magnetized pulsars. 
 
\subsubsection{PSR 1509-58 and high B pulsars} 
 \label{sec:highB} 
 
Having established the generic properties of the more familiar 
gamma-ray pulsars, it is salient to move on to their more 
highly-magnetized siblings.  The operative cascade physics is 
identified in Section~2.  High fields of pulsars like PSR 1509-58 
inhibit cascading (BH01) via the suppression of pair creation.  This 
removes the complexity of the sequence of synchrotron components and 
leaves a bare curvature (or inverse Compton) primary electron spectrum, 
which is generically flat.  The spectral index of PSR 1509-58, roughly 
1.6 (e.g.  see HBG97), is consistent with curvature emission from 
cooled primaries.  Accordingly, it can be deduced that increasing the 
surface field must at some point flatten the pulsar continuum.  This 
starkly contrasts the inference from Eq.~(\ref{eq:alpha_n}) of 
steepening with increasing \teq{B_0}; this trend reversal, probably at 
around \teq{B_0\sim 10^{13}}Gauss, is a distinctive prediction of the 
polar cap model.  Therefore, to address the focus of this paper, based 
on these spectral slope issues, it appears unlikely that {\it on-beam} 
high field pulsars are optimal candidates for EGRET UIDs, which appear 
to have spectra steeper on average than \teq{\erg^{-2}}, at least those 
correlated with the Gould Belt (Gehrels et al. 2000, Grenier et al. 
2000; see also the reviews by Grenier and Gehrels in these 
proceedings).  Note that the determination of UID spectral indices is 
subject to selection effects against detecting steep spectrum sources; 
this suggests that the true average may be even steeper. 
 
This pessimism is reinforced by the expectation that high field pulsars 
have low maximum energies, according to the pair creation turnovers 
predicted and observed in Figure~\ref{fig:hecutoff}.  This is 
dramatically emphasized by the lack of an EGRET detection for PSR 
1509-58.  Such an absence of emission is extremely constraining on 
spectral models, and was exploited by HBG97 to assert that photon 
splitting was acting in PSR 1509-58.  Splitting attenuates photons at 
somewhat lower energies than does pair creation in such high B pulsars 
(see BH01 for details).  This implies that the true \teq{\emax} curves 
in Figure~\ref{fig:hecutoff} lie somewhat lower than the \teq{\gamma\to 
e^{\pm}} ones displayed at \teq{B\gtrsim 10^{13}}Gauss when 
\teq{\gamma\to\gamma\gamma} is taken into account.  HBG97 observed that 
splitting naturally accounts (pair creation alone cannot, as is obvious 
from the figure) for the inferred turnover at around 10--30 MeV in PSR 
1509-58 if a standard polar cap size is assumed and general relativity 
is incorporated in photon transport calculations.  This study provided 
a nice piece of circumstantial evidence for the action of photon 
splitting, an interesting prospect for physics. 
 
Returning to the EGRET UIDs, these opacity constraints indicate that 
high B sources can only be candidates if their emission regions are at 
relatively high altitudes, for which there is no confirmed 
observational evidence.  Nevertheless, if the galactic UIDs turn out to 
be high B pulsars, they would then be expected to have steep spectra if 
the cutoff always matches the EGRET band, an improbable fine-tuning. 
The cutoffs would not be super-exponential, the signature of pair 
creation, but rather more gradual if splitting is operating.  In such a 
case, there would be a segregation of flatter Crab-like and Vela-like 
pulsars in the galactic plane, and steep-spectrum higher field pulsars 
at low to moderate galactic latitudes.  How such a correlation between 
\teq{B_0} and latitude (i.e.  perhaps also kick velocity) would be 
attained is presently unclear. 
 
Before closing this subsection it is desirable to make a small pitch 
for {\bf gamma-ray polarimetry}.  This is perhaps most relevant to high 
B pulsars, but is still quite important for the more common EGRET 
pulsars.  Gamma-ray polarimetry is traditionally a haven for skeptics, 
though the mood of the high energy astrophysics community is rapidly 
changing given the prospects (Lei, Dean \& Hills, 1997) that the 
INTEGRAL mission will detect polarization at the 10\% level from the 
Crab pulsar (at 200--600 keV), and also in a handful of other sources. 
Hard gamma-ray experiments like GLAST are generally not afforded the 
opportunity to act as polarimeters, being limited by multiple 
scattering in trackers above 300 MeV.  Despite early estimates of 
GLAST's potential polarimetric capability (Yadigaroglu 1997), current 
design precludes a major focus on this observational goal.  Medium 
energy gamma-ray experiments, on the other hand, are ideally suited to 
polarization studies (via their sampling of Compton scattering 
kinematics), and accordingly considerable emphasis has been placed in 
recent workshops on such new developments for next-generation advanced 
Compton telescopes (e.g. see Kanbach et al. 2000, and the Web pages for 
the MEGA [{\tt http://www.gamma.mpe-garching.mpg.de/MEGA/mega.html}] 
and ACT [{\tt http://gamma.nrl.navy.mil/ngram/}] consortia).  Science 
motivations for polarimetry are obvious for pulsars.  The presence of 
strong fields virtually guarantees a strong polarization signal in 
polar cap models, and when these couple with spectral structure and 
temporal information, particularly powerful observational diagnostics 
are achieved.  This may be fruitful at the lower end of the cascade 
continuum in Vela-like objects, but it is an especially valuable tool 
for highly-magnetized pulsars since the attenuation cutoffs fall in the 
Comptel band, and should exhibit strong and distinctive polarization 
signatures.  A concerted effort to realize the historically ambitious 
goal of gamma-ray polarimetry may yield dramatic science gains in the 
near future. 
 
\subsubsection{Magnetars: not relevant for UIDs?} 
 \label{sec:magnetar} 
 
A natural step from these considerations of high field pulsars is to 
magnetars, specifically anomalous X-ray pulsars (AXPs) and soft gamma 
repeaters (SGRs).  These can be quickly dismissed as potential 
candidates for EGRET UIDs unless we have been unlucky in the 
observational process.  To date, identified emission from these sources 
does not exceed 1 MeV for the SGRs (e.g. see data for giant flares in 
Mazets et al. 1981 for the 5th March 1979 [SGR 0525-66] event, and 
Hurley et al 1999 for SGR 1900+14) and considerably less for AXPs. 
Moreover, their spectra are typically steep, at least in quiescent 
epochs, generally precluding detection in the EGRET band.  The possible 
exceptions are the so-called ``initial hard spikes'' in the two cases 
of giant flares of SGR 0525-66 (in 1979) and SGR 1900+14 (in 1998), 
neither of which was seen by a hard gamma-ray mission.  Hence, 
observationally there is still the possibility that this particular 
mode of activity may generate super-MeV emission, a question that GLAST 
might be able to answer.  From the theoretical viewpoint, high energy 
emission is not readily expected from these sources, if the magnetar 
interpretation is adopted.  This is because the magnetospheric opacity 
due to pair creation and photon splitting will inhibit escape of 
photons above around 50--200 MeV (e.g. see Figure~9 of BH01).  This 
bound applies specifically to a ``pulsar mode,'' where (e.g. giant 
flare) emission is strongly coupled to a beam near the polar cap; 
extending to equatorial regions enhances the opacity and can force the 
spectrum down into the hard X-ray/soft gamma-ray band (Baring 1995, 
Harding \& Baring 1997), more commensurate with that seen in normal SGR 
outbursts and AXPs.  The escape clause here is to reduce the strength 
of the ambient magnetic field, i.e. either to relinquish (untenable to 
some) the magnetar interpretation, or to move the giant flare emission 
region to higher altitudes (not necessarily outer gaps), a prospect 
that is difficult to reconcile with the enormous energy liberated in 
SGR giant flares. 
 
\subsubsection{Radio quiescence at high B?} 
 \label{sec:quiescence} 
 
An issue that impacts the discussion of unidentified gamma-ray sources 
is the existence or otherwise of radio counterparts.  This concerns 
highly-magnetized pulsars, if they can be radio quiet without 
dramatically inhibiting gamma-ray emission, as has been suggested by 
Baring and Harding (1998, see also Zhang \& Harding 2000b).  Since it 
is commonly assumed  (e.g.  Sturrock 1971; Ruderman \& Sutherland 1975; 
Arons \& Scharlemann 1979; for a dissenting view, see Weatherall \& 
Eilek 1997) that a plentiful supply of pairs is a prerequisite for 
coherent radio emission at observable flux levels, any suppression of 
pair creation in pulsars implies that the emission of radio waves 
should be strongly inhibited.  Baring \& Harding (1998) posited the 
idea that photon splitting could effect a suppression of pair creation 
by providing a more competitive mode of photon attenuation for high 
polar fields.  Accordingly, they predicted an approximate boundary in 
the \teq{P}-\teq{{\dot P}} diagram that delineated pulsars of lower 
\teq{{\dot P}} (or \teq{B_0}) that could be radio-loud, and those of 
unusually large period derivative, which where necessarily radio quiet 
due to the action of splitting.  The boundary was: 
\begin{equation} 
   \dot{P}\;\approx\; 7.9\times 10^{-13}\;  
   \biggl(\dover{P}{1\,\hbox{sec}}\biggr)^{-11/15}\quad . 
 \label{eq:deathline} 
\end{equation} 
The fact that this {\it boundary of quiescence} neatly separated 
members of the 1995 version of the Princeton Pulsar Catalog (Taylor et 
al. 1993) from the small family of purportedly radio-quiet magnetars 
(i.e. AXPs and SGRs), was an enticing piece of support for the 
proposition.  Yet this concept pre-dated results from the new Parkes 
Multi-Beam survey [{\tt 
http://www.atnf.csiro.au/\~{}pulsar/psr/pmsurv/pmwww/}] that has 
\linebreak discovered a small number of pulsars of higher magnetization 
than previously known (e.g. Camilo et al. 2000), with three lying above 
this putative quiescence boundary (BH01).  This development proves not 
to be unduly disturbing, since only a small change in the emission 
altitude can comfortably accommodate the new detections (BH01). 
Moreover, the ``polarity'' of the rotating magnetosphere can influence 
the nature of the acceleration gap, with significant consequences for 
the boundary of quiescence (Zhang \& Harding 2000b).  What is more 
telling, from an observational perspective, is that one of the radio 
pulsars recently discovered in the Parkes Multi-Beam Survey, PSR 
J1814-1744, lies very close to the anomalous X-ray (AXP) pulsar CTB 
109, so that a single radio quiescence line cannot separate the radio 
pulsar and AXP/magnetar populations.  This proximity coupled with the 
fact that PSR J1814-1744 has not been detected in X-rays (Pivovaroff, 
Kaspi \& Camilo 2000) strongly suggests that a quantity other than 
\teq{P} and \teq{B_0} has a profound influence on the properties of 
highly-magnetized radio pulsars and AXPs. 
 
From a theoretical standpoint, the suppression of pair creation at high 
fields by photon splitting is not unequivocal.  The extensive 
investigation by Baring \& Harding (2001) of photon propagation and 
attenuation in general relativistic magnetospheres revealed that 
suppression was significant only if both polarization states (in the 
external magnetic field) of photons could split, rather than just one. 
This point addresses a subtlety of QED dispersion of the magnetized 
vacuum, principally in relation to selection rules derived by Adler 
(1971): only one polarization state is amenable to splitting in the 
limit of weak to moderate vacuum dispersion.  While almost certainly 
applicable to typical pulsars, this contention, based on the leading 
order contribution to the dispersive properties provided by the vacuum, 
may or may not persist in supercritical (\teq{B\gtrsim 4.41\times 
10^{13}}Gauss) fields where higher order QED corrections become 
operative.  Hence, whether or not splitting can act to inhibit pair 
creation critically depends on this unanswered question of physics, the 
mathematical solution of which is potentially difficult or 
intractable. 
 
\subsubsection{Off-Beam Pulsars} 
 \label{sec:offbeam} 
 
The situation concerning on-beam pulsars motivates an expansion of
perspective.  From the foregoing discussions, EGRET-type pulsars
typically have spectra flatter than EGRET UIDs, and high field
counterparts might match the UID slopes if a conspiracy establishes
their turnovers at just the right energies to mimic the steeper UID
spectra.  In the absence of a comfortable explanation of UIDs within
the context of on-beam polar cap pulsars, Harding \& Zhang (2001)
recently proposed {\it off-beam} pulsars as candidates for some UIDs
with Gould Belt associations.  Effectively, the line of sight to Earth
does not cut the rim of the polar cap in these sources, but rather
samples a broader (spatial) wing corresponding to higher altitudes
above the neutron star surface, from which the emission is typically at
lower energies, but with a harder spectrum.  The hardness originates in
the curvature primary photons, with a simultaneous drop in the maximum
energy due to the combination of pair creation attenuation and field
geometry.  The net effect is that the spectrum in the EGRET band is
steeper for these sources, due to the influence of a cutoff in the
near-GeV range, however the solid angle of emission increases from that
of on-beam pulsars.  Hence, pulsation searches will be biased towards
on-beam pulsars despite off-beam ones constituting a larger percentage
of the population.  While an attractive proposition in several ways,
this suggestion still mandates some fine-tuning of the observational
perspective to generate spectra that match UID observations.  This
issue plagues the high-field pulsar explanation also, and in fact,
these two alternatives pose a challenge: how can one discriminate
between off-beam/moderate B and on-beam/high B scenarios given that
they display similar spectral properties.  The answer may be provided
by population statistics.
 
\section{Global Properties for Population Studies} 
 
\subsubsection{Gamma-Ray Luminosities} 
 
As indicated in the Introduction, one of the principal successes of the 
polar cap model is its prediction (Harding 1981) of an almost linear 
correlation  between the inferred luminosity of gamma-ray pulsars and 
\teq{B_0/P^2}, i.e. the voltage across the open field lines for 
standard polar caps.  This correlation, while not exact, largely due to 
the uncertainty in determining source distance by folding (radio) 
dispersion measures into the Taylor-Cordes (1993) galactic electron 
model, is distinctly different from the canonical spin-down luminosity, 
\teq{\propto B_0^2/P^4}, which is mirrored by the X-ray pulsar 
population (Becker \& Tr\"umper 1997).  An enticing feature of this 
prediction was that only 2 gamma-ray pulsars were known at the time it 
was proposed, and subsequent predictions by competing 
analyses/models (e.g. Sturner \& Dermer 1994; Romani \& Yadigaroglu 
1995; Cheng \& Zhang 1998; Rudak \& Dyks 1999) and revisions (Zhang \& 
Harding 2000a) all post-dated the EGRET database.  The current status is 
that the polar cap expectations (Sturner \& Dermer 1994; Zhang \& 
Harding 2000a) match the data more accurately than their outer gap 
counterparts (Romani \& Yadigaroglu 1995; Cheng \& Zhang 1998), with 
each group of researchers offering different \teq{B_0} and \teq{P} 
dependences for the luminosity (see Harding 2001a for a review).   To 
some extent, this situation is limited by small number statistics, an 
issue that will be irrelevant in the GLAST era, when such correlations 
will be established on a really firm basis. 
 
Setting aside partisan theoretical justifications, this 
observational correlation motivates a revision of historical thinking. 
Traditionally, the EGRET community has used the spin-down luminosity 
\teq{B_0^2/P^4} as an indicator of a pulsar's observability.  While 
theoretically motivated in some sense, this choice does not match the 
established trend, and can dictate periods that are selected in 
pulsation searches.  While this has netted most pulsars high up on a 
\teq{B_0^2/P^4/\dpsrsq} rank-ordered list (where \teq{\dpsr} is the 
pulsar distance), certain gamma-ray pulsars (notably the longer period 
pulsars PSR 0656+14 and PSR 1055-52) are surprisingly low in spin-down 
luminosity, and millisecond pulsars have proven extraordinarily 
difficult to detect (up till PSR 0218+4232, see Kuiper et al. 2000) 
given their short periods.  Clearly, a gamma-ray luminosity dependence 
\teq{L_{\gamma}(P,\, {\dot P})} that differs from the spin-down one 
will dramatically modify the observability criterion, particularly if 
the period dependence is substantially different.  Furthermore, the 
spectral shape also influences the observability (Baring \& Harding 
2000), a more subtle influence.  This is a consequence of how the 
luminosity is distributed in the gamma-ray band, specifically that 
portion that emerges above the threshold sensitivity for a specific 
gamma-ray detector. The driving parameters for such an apportionment 
are \teq{\emax}, index \teq{\alpha_n}, and to a lesser extent \teq{\emin}, 
since the spectra are generally flat enough for the bulk of the 
luminosity to emerge at the highest energies.  These parameters control 
the normalization of the pulsar gamma-ray power-law. 
  
An appropriate definition of a detector's observability \teq{{\cal 
O}(\ethresh )} is the {\it integral} flux above an effective 
instrumental energy threshold \teq{\ethresh}.  For pulsars with 
\teq{\emin\ll\ethresh}, the usual case for GLAST considerations, this 
scales as the luminosity divided by the spectral normalization, 
yielding \teq{{\cal O}(\ethresh )\propto L_{\gamma}(P,\, {\dot P})\, 
\emax^{(\alpha_n-2)}/\dpsrsq} (Baring \& Harding 2000).  Modifications 
to this dependence are possible, in particular if \teq{\emin\gtrsim 
\ethresh}, in which case \teq{{\cal O}(\ethresh )\propto 
L_{\gamma}(P,\, {\dot P})/\emin/\dpsrsq}.  Either of these 
possibilities yields substantially different observabilities from the 
spin-down formula, assuming that \teq{\emax} and \teq{\emin} scale with 
\teq{B_0} and \teq{P} approximately as the low field alternative 
offered in Eq.~(\ref{eq:emax}).  This leads to the conclusion that 
observabilities predicted for GLAST pulsation searches should follow a 
dependence somewhere in between \teq{B_0/P^2} and \teq{B_0^2/P^{5/2}}. 
Using the latter possibility, Baring \& Harding (2000) generated a 
revised rank-ordered listing that indicated a dramatic rearrangement 
from the traditional EGRET ordering.  Notable changes included the 
much higher ranking of the ``outlier'' longer period gamma-ray pulsars 
PSR 0656+14 and PSR 1055-52, and the marked lowering of millisecond 
pulsars (PSR 1939+2134, PSR 0437-4715, PSR 1744-1134, etc.) in the 
ranks, specifically out of the top 40.  Both of these reflect the 
weaker dependence of the revised observability on \teq{P}.  The old and 
new rankings are listed in Table~1 for the confirmed and candidate (non-millisecond) 
gamma-ray pulsars; it becomes clear that such revisions mute 
questions of why PSR 1055-52 was seen by EGRET.  Refinements of such 
rank orderings are in progress, an interesting preparatory step for the 
GLAST mission. 
 
\begin{table}[ht] 
\caption[CGRO Gamma-Ray Pulsars and rank-ordered listings] 
{CGRO Gamma-Ray Pulsars and rank-ordered listings} 
\begin{tabular*}{\textwidth}{@{\extracolsep{\fill}}lcccc} 
\hline 
\it PSR & \it $P$ (sec) & ${\dot P}$ &$B_0^2/P^4/\dpsrsq$ & 
$B_0^2/P^{5/2}/\dpsrsq$\cr 
&&&\it rank&\it rank\cr 
\hline 
Vela    & 0.089 & $1.25\times 10^{-13}$ & 2 & 1\cr 
Crab    & 0.033 & $4.21\times 10^{-13}$ & 1 & 2\cr 
Geminga & 0.237 & $1.1\times 10^{-14}$ & 4 & 3\cr 
1509-58 & 0.150 & $1.5\times 10^{-12}$ & 5 & 5\cr 
1706-44 & 0.102 & $9.3\times 10^{-14}$ & 7 & 6\cr 
0656+14 & 0.385 & $5.5\times 10^{-14}$ & 20 & 13\cr 
1951+32 & 0.040 & $5.85\times 10^{-15}$ & 6 & 14\cr 
1055-52 & 0.197 & $5.83\times 10^{-15}$ & 33 & 23\cr 
\hline 
\end{tabular*} 
%
\end{table} 
\inxx{captions,table}

\subsubsection{Gamma-Ray vs. Radio Observability} 
  
Pulsation searches for EGRET UIDs that have spatial associations with 
known radio pulsars naturally bias the search phase space in \teq{P} 
and \teq{\dot P}.  Yet there is no guarantee that every gamma-ray 
pulsar (or EGRET UID) is a radio pulsar, and vice versa.  The polar cap 
and outer gap models make distinctly different predictions of the 
correlation between observing gamma-ray and radio emission from 
pulsars, based largely on assumed emission region geometries.  Outer 
gap models (e.g.  Romani \& Yadigaroglu 1995; Zhang et al. 2000) 
suggest an almost complete disconnect between emission in the two 
wavebands so that detected (normal as opposed to high B) gamma-ray 
pulsars should mostly be radio-quiet due to the much larger solid angle 
in the gamma-ray beam.  The fact that Geminga is the only radio-quiet 
gamma-ray pulsar may be bothersome to outer gap proponents.  Perhaps 
more disconcerting is that the best determinations of altitudes for the 
radio emission (e.g. Gil \& Han 1996) may suggest more of a connection 
with polar caps than outer gaps.  Even if the origin of both radio and 
gamma-ray emission is connected to pair creation in a polar cap, the 
solid angles of these components should be somewhat different.  There 
are large uncertainties present in any prediction of the ratio of 
numbers of gamma-ray and radio pulsars, since models must incorporate 
details of the distribution of fields and periods at birth, the spatial 
and velocity distributions of pulsars in the galaxy, the influence of 
galactic gravitational potentials, and luminosity and solid angle 
geometry prescriptions for both the gamma-ray and radio emission. 
Assembling such ingredients, both Sturner and Dermer (1996) and the 
very recent analysis of Gonthier et al. (2001) find that only a 
minority of EGRET pulsars would be expected to be radio-quiet.  This 
fraction increases substantially for GLAST to an almost 50/50 
radio-quiet/radio-loud situation according to Gonthier et al. (2001); 
this result is a consequence of GLAST's improved sensitivity enabling 
it to sample deeper than typical radio surveys.  Hence blind pulsation 
searches shall be a much more salient tool for GLAST, and the gamma-ray 
UID community may well have to forgo attachments to radio pulsar 
counterparts. 
 
\section{Conclusion} 
 
This review has detailed some of the expectations for polar cap pulsar 
properties that will prove extremely useful subsequent to pulsar 
identification for gamma-ray sources.  These include spectral trends 
with \teq{B_0} and \teq{P}, identifying the science gains to 
be made given a considerable gamma-ray pulsar database in 
the GLAST era.  Yet the discussion has also elucidated possible 
guidelines for UID pulsation searches so that pulsar science is not 
merely a {\it post-facto} consideration for the identification 
process.  It is clear that standard gamma-ray pulsars like the Crab, 
Geminga and Vela might not constitute the majority of the subset of 
UIDs that eventually turn out to be pulsars, at least if the mean GLAST 
gamma-ray source characteristics are similar to those of the EGRET 
UIDs.  Turning to off-beam pulsars or highly-magnetized on-beam ones as 
candidates does not dramatically change the search phase space (\teq{P} 
and \teq{\dot P}) from current agendas.  Lowering the instrumental 
threshold energy as much as possible is a worthwhile goal, provided 
that it does not compromise threshold sensitivity and angular 
resolution properties.  While it has been argued that observability 
criteria need revision from historical preferences, and that there is 
probably less need to be biased against searching at longer periods, 
since magnetars appear to be unlikely candidates for UIDs, there is no 
compelling reason to search on supersecond periods and in the domain 
\teq{{\dot P}\gtrsim 3\times 10^{-12}} sec/sec.  Continued identification 
efforts with the EGRET database and refinements to theoretical models 
in the next half decade will help set the stage for the watershed of  
gamma-ray identifications to be attained by GLAST.


\end{document}